\documentclass[twocolumn,showpacs,preprintnumbers,pra,aps,amssymb,amsfonts,amstex]{revtex4}

\usepackage{times,graphicx}

\newcommand{\trace}{\mathop{\rm Tr}\nolimits}

\newcommand{\diag}{\mathop{\rm Diag}\nolimits}

\newcommand{\eig}{\mathop{\rm Eig}\nolimits}

\newcommand{\bra}[1]{\langle#1|}
\newcommand{\ket}[1]{|#1\rangle}

\newcommand{\qed}{\hfill$\square$\par\vskip24pt}

\newcommand{\be}{\begin{equation}}
\newcommand{\ee}{\end{equation}}
\newcommand{\bea}{\begin{eqnarray}}
\newcommand{\eea}{\end{eqnarray}}
\newcommand{\beas}{\begin{eqnarray*}}
\newcommand{\eeas}{\end{eqnarray*}}

\newcommand{\id}{\mathrm{\openone}}

\newtheorem{theorem}{Theorem}

\begin{document}
\title{A Sharp Fannes-type Inequality for the von Neumann Entropy}
\author{Koenraad M.R. Audenaert}
\email{k.audenaert@imperial.ac.uk}
\affiliation{Institute for Mathematical Sciences, Imperial College London,
53 Princes Gate, London SW7 2PG, UK}

\date{\today}

\begin{abstract}
We derive an inequality relating the entropy difference between two quantum states
to their trace norm distance, sharpening a well-known inequality due to M.\ Fannes.
In our inequality, equality can be attained for every prescribed value of the trace norm distance.
\end{abstract}
\pacs{03.65.Hk}
\maketitle
\section{Introduction}
The initial motivation of the present paper was given in by a purely pedagogical issue:
given the ubiquity of powerful computers on nearly every desk \cite{microsoft},
one should be able to quickly illustrate (rather than prove) the validity of many basic inequalities.
In quantum mechanics, and in Quantum Information Theory in particular, perhaps the best known
inequality is the eponymous continuity inequality (\ref{eq:fannes1})
for the von Neumann entropy, discovered by M.\ Fannes.
This inequality gives an upper bound on the absolute value of the difference between the von Neumann entropies
of two finite-dimensional quantum states, in terms of their trace norm distance (\ref{eq:Tdef}).

The inequality can easily be illustrated using a computer, as it deals with finite-dimensional
quantum states and each of its constituents can be calculated efficiently.
What one has to do is to generate random pairs of states, calculate both the trace norm distance
and the absolute value of the difference of their von Neumann entropies, and produce a scatter plot of these two
quantities. Adding to that a graph of the upper bound, one should see a cloud of points lying
below the latter graph. Indeed, only a few minutes of work is required to produce plots
akin to those of Figures 1, 2 and 3 \cite{provided}.

Now one directly sees that the bound is indeed an upper bound, but also that the bound is not sharp.
There are no points on the graph, or even near it. Although this is certainly not a problem for the originally intended use
of the bound -- proving a continuity property of the von Neumann entropy --
 nevertheless, like the present author, one could be compelled to find a better bound; a sharp bound,
that exactly describes the upper boundary of the cloud of randomly generated points.

In \cite{baka2}, the author, together with J.~Eisert, did exactly this for the relative entropy,
which is in a sense a quantity derived from the von Neumann entropy.
In the present paper, the same is done for the von Neumann entropy itself.
The present paper could therefore be considered the `prequel' of \cite{baka2}.
The outcome is a new, sharp bound, of the same type as Fannes' one, and, rather surprisingly,
of the same complexity.

As mentioned, there are no real benefits in the new bound w.r.t.\ proving continuity of the von Neumann entropy.
However, in recent times, new usage of such a bound has been found, e.g.\ in entanglement theory. For this modern usage
our bound has the important benefit that it is actually easier to use, because it is valid over the whole range of possible values
of the trace norm distance, unlike Fannes' one, which only holds for trace norm distances less than $1/e$ and has to
be modified for larger ones.
Furthermore, it is the sharpest bound possible and improves on the older one.
The only added cost of the new bound goes in its proof, which is much longer.

\bigskip

Before stating the main result, let us first introduce some notations.
The acronyms LHS and RHS are short for left-hand side and right-hand side.
To denote Hermitian conjugate, we follow mathematical conventions and use the asterisk rather than the dagger.
The notation $\diag(x,y,z\ldots)$ denotes the diagonal matrix with diagonal elements $x,y,z,\ldots$, and
$\eig^\downarrow(A)$ denotes the vector of eigenvalues of a Hermitian matrix $A$, sorted in non-increasing order.
Following information-theoretical convention, we use base-2 logarithms, denoted by $\log_2$.
The natural logarithm will be denoted by $\ln$.
The von Neumann (vN) entropy, when expressed in units of qubits, is then defined as
\be
S(\rho):=-\trace[\rho\log_2\rho].
\ee
For classical probability distributions, this reduces to the Shannon entropy
\be
H(p):= -\sum_i p_i\log_2 p_i,
\ee
where $p$ is a probability vector.
We will occasionally indulge in overloaded usage of the symbol $H$ and define $H(x):= -x\log_2 x$ for non-negative scalars $x$.
Thus the relation $H(p)=\sum_i H(p_i)$ holds.

We use the following definition for trace norm distance:
\be\label{eq:Tdef}
T(\rho,\sigma)=||\rho-\sigma||_1/2,
\ee
including the factor $1/2$ to have $T$ between 0 and 1.

\bigskip

The original inequality for the continuity of the vN entropy, as proven by Fannes \cite{fannes,mikeandike},
reads:
\be\label{eq:fannes1}
|S(\rho)-S(\sigma)| \le 2T\log_2(d)-2T\log_2(2T),
\ee
which is valid for $0\le T\le 1/2e$.
For larger $T$ one can use the weaker inequality
\be\label{eq:fannes2}
|S(\rho)-S(\sigma)| \le 2T\log_2(d)+1/(e\ln(2)).
\ee

Our main result is a sharpening of these inequalities:
\begin{theorem}
For all $d$-dimensional states $\rho$, $\sigma$ such that their trace norm distance
is given by $T$,
\be\label{eq:sharp}
|S(\rho)-S(\sigma)| \le T\log_2(d-1) + H((T,1-T)).
\ee
\end{theorem}
In fact, by construction of this bound, there is no sharper bound than this one
that exploits knowledge of $T$ and $d$ only.

To show that sharpness holds for any value of $T$ and $d$, we just note that the following pair of
(commuting) states achieves the bound:
\bea
\rho   &=& \diag(1-T,T/(d-1),\ldots,T/(d-1)) \\
\sigma &=& \diag(1,0,\ldots,0).
\eea
In other notations:
\bea
\sigma &=& \ket{0}\bra{0} \\
\rho &=& \frac{Td}{d-1}\,\, \frac{\id_d}{d} + \left(1-\frac{Td}{d-1}\right)\ket{0}\bra{0}.
\eea
Note that the coefficient of $\ket{0}\bra{0}$ in $\rho$ may be negative.
A simple calculation then yields that their trace norm distance is $T$, and their entropy difference is
$T\log_2(d-1) + H((T,1-T))$. We once again stress that Fannes' original bound is not sharp: there are no pairs of states
saturating Fannes' bound except in the trivial case when they are identical ($T=0$).

\begin{figure}
\includegraphics[width=3.3in]{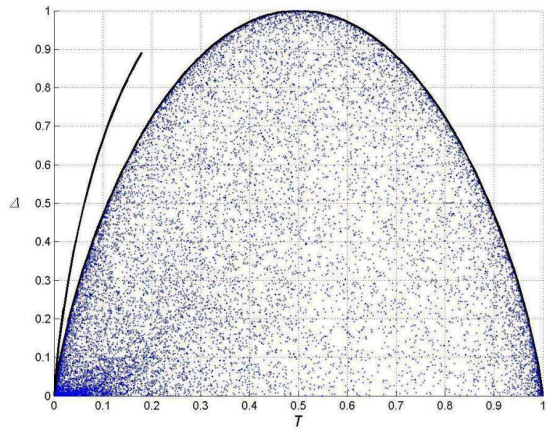}
\caption{\label{fig1}
Scatter plot of 20000 randomly generated pairs $(\rho,\sigma)$ of qubit states ($d=2$);
shown is the trace norm distance $T=||\rho-\sigma||_1/2$ versus the difference $\Delta=|S(\rho)-S(\sigma)|$
of the vN entropies.
The upper curve in the interval $0\le T\le 1/(2e)$
represents the Fannes bound (\ref{eq:fannes1}). The lower curve represents our sharp bound (\ref{eq:sharp})
and is seen to follow the boundary of the set of scatter points tightly.
}
\end{figure}

\begin{figure}
\includegraphics[width=3.3in]{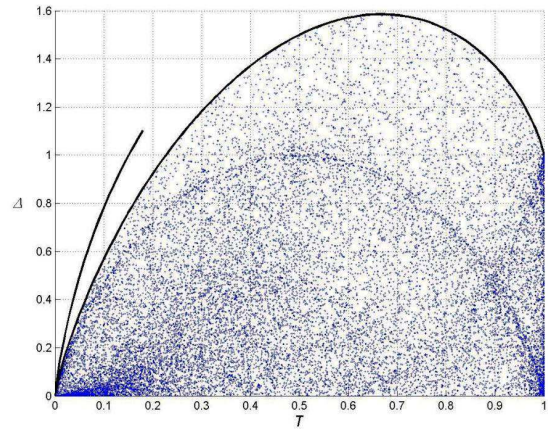}
\caption{\label{fig2}
Same as Fig.\ 1, but for qutrits ($d=3$).
}
\end{figure}

\begin{figure}
\includegraphics[width=3.3in]{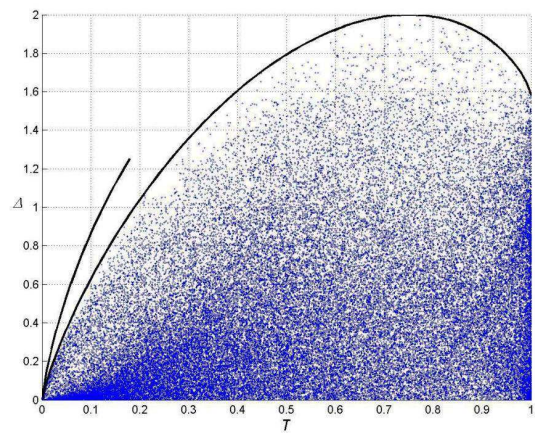}
\caption{\label{fig3}
Same as Fig.\ 1, but for 4-dimensional quantum systems ($d=4$).
}
\end{figure}
\section{Proof}
The remainder of this paper will be devoted to the proof of our inequality.
Because of its complexity, we will proceed in several stages.
\subsection{Reduction to classical case}
The first step of the proof is to reduce the statement to the commuting (classical) case.
Since $S$ is unitarily invariant, $S(\rho)$ only depends on the eigenvalues of $\rho$.
Let us denote the eigenvalue decompositions of $\rho$ and $\sigma$ by $\rho=V\diag(\Lambda_\rho) V^*$ and
$\sigma=W\diag(\Lambda_\sigma) W^*$; here, $\Lambda_\rho = \eig^\downarrow(\rho)$.
The LHS of (\ref{eq:sharp}) then becomes $|H(\Lambda_\rho) - H(\Lambda_\sigma)|$, and the trace norm distance, which is
the only ingredient of the RHS that depends on the states, is given by
$T=||\diag(\Lambda_\rho)-U\diag(\Lambda_\sigma)U^*||_1/2$, where $U=V^* W$.

Let us now fix the eigenvalues of $\rho$ and $\sigma$; the only degree of freedom is then in the unitary matrix $U$, which
only appears in the RHS. The LHS is thus fixed, while the RHS can be varied. Referring to the Figures, this amounts to looking
at cross-sections of the plot along the horizontal lines. To prove correctness of the bound (\ref{eq:sharp}) we have to look at the points
of minimal (leftmost) and maximal (rightmost) trace norm distance.
Inequality IV.62 in \cite{bhatia}, which essentially seems to be due to Mirsky \cite{mirsky}, reads:
\beas
|||\eig^\downarrow(A)-\eig^\downarrow(B)||| &\le& |||A-B||| \\
|||A-B||| &\le& |||\eig^\downarrow(A)-\eig^\uparrow(B)|||,
\eeas
for all Hermitian $A$ and $B$ and all unitarily invariant norms.
In particular, we get that the extremal values of $T=||\diag(\Lambda_\rho)-U\diag(\Lambda_\sigma)U^*||_1/2$, when varying
$U$, are obtained for $U$ equal to certain permutation matrices. More precisely, the minimal value
is obtained for $U=\id$, and the maximal value for $U$ the permutation matrix that totally reverses the diagonal entries.

This shows, in particular, that the boundary of the ``point cloud'' can be found for diagonal $\rho$ and $\sigma$, i.e.\ for
commuting states.
\subsection{Proof Strategy}
In the following we can therefore restrict to the commuting case and only look at (discrete) probability distributions
and their Shannon entropies.
To highlight the classical nature of the remainder of the proof, we will replace the states $\rho$ and $\sigma$ by
$d$-dimensional probability vectors $p$ and $q$. We have to show that the following inequality holds:
\be\label{eq:H}
|H(p)-H(q)| \le T\log_2(d-1) + H((T,1-T)),
\ee
where $T$ is now
\be\label{eq:T}
T:=(1/2)\sum_{i=1}^d |p_i-q_i|.
\ee
We will do this in a constructive way, by fixing $T$ and looking for pairs $p,q$ that maximise the LHS.
The maximal value of the LHS thus obtained then will be a sharp upper bound by construction.

At this point it is interesting to mention that simple things don't work. For example,
it is not obvious that $|H(p)-H(q)|$ should be maximal for $p$
``pure'', because this quantity is neither convex nor concave, and furthermore is to be maximised
over the rather complicated set of all $(p,q)$
such that $(1/2)\sum_{i=1}^d |p_i-q_i| = T$, $p_i\ge0$, $q_i\ge0$ and
$\sum_i p_i=\sum_i q_i=1$ hold.

Let us introduce the symbol $\delta:=p-q$. Since $p$ and $q$ are probability vectors, the
$\delta_i$ are real numbers adding up to 0. We can decompose $\delta$ in a positive and negative part, which we
denote by $\delta^+$ and $\delta^-$. Thus we have $\delta=\delta^+ - \delta^-$.
Both parts consist of non-negative reals and their elementwise product $\delta^+_i \delta^-_i$
is 0.
The constraint (\ref{eq:T}) then translates to $\sum_i \delta^+_i = T$ and $\sum_i \delta^-_i = T$.

In the following, we will shift attention to the quantity $H(q)-H(p)$ (without taking absolute values)
and try to find its global minimum. Subsequently taking the absolute value then yields the maximum
of $|H(q)-H(p)|$.

\subsection{The case $d=2$}
When $d=2$, we automatically get that $\delta$ must be given by $\delta=(+T,-T)$.
The quantity to be minimised is then
$$
H(q)-H(p) = H((p_1+T,1-p_1-T))-H((p_1,1-p_1)),
$$
where $p_1$ is the first entry of $p$.
As this quantity is obtained by setting $d=2$ in (\ref{eq:v2}) below, we need not spend more time on this
special case. The reader is advised to proceed to the end of subsection G and thereby collect a free parking token \cite{attention}.
\subsection{Optimal $\delta^+$}
We will prove here that the optimal $\delta^+$ is ``rank 1''; that is, it has just one non-zero entry, which then is given by $T$.
W.l.o.g., since nothing has been claimed yet about $p$ or $q$ themselves, we can put this non-zero entry on the first position.
Furthermore, $\delta^-$ can then take non-zero values on all positions except the first one.

Letting $p_1$ be the first entry of $p$, $p$ and $q$ must then be of the form
\bea
p &=& (p_1,(1-p_1)r) \\
q &=& (p_1+T,(1-p_1)r-Ts),
\eea
where $r$ and $s$ are $(d-1)$-dimensional probability vectors, with the restrictions
\bea
p_1+T &\le& 1 \\
(1-p_1)r-Ts &\ge& 0.
\eea
Here, $Ts$ is just $\delta^-$.
The value of $H(q)-H(p)$ corresponding to this is given by
\bea
H(q)-H(p) &=& H(p_1+T)-H(p_1) \nonumber\\
&& + H((1-p_1)r-Ts) \nonumber\\
&& - H((1-p_1)r). \label{eq:hqhp}
\eea
The remaining minimisation over $r$, $s$ and $p_1$ will be performed in the subsequent stages.

\bigskip

\textit{Proof.}
Let us now prove that the optimal $\delta^+$ must indeed be rank 1.
So we put $q=p+\delta^+-\delta^-$ and fix $p$ and $\delta^-$, under the restrictions $p-\delta^-\ge0$.
The restrictions on $\delta^+$ are, as mentioned before,
$\delta^+\ge0$, $\delta^+\delta^-=0$, and $\sum_i \delta^+_i = T$.
Hence, $\delta^+$ is restricted to a convex set. If $\delta^-$ is 0 on the positions 1 to $k$, say, then the extremal points of
this convex set are given by $Te^1,Te^2,\ldots,Te^k$.
Now the optimality of one of these extremal points follows because
$H(q)-H(p) = H(p+\delta^+-\delta^-)-H(p)$ is concave in $\delta^+$ (since $H$ is concave), and it is well-known that concave functions
reach their global minimum over a convex set in one (or more) of the extremal points of that set.
\qed
\subsection{Optimal $(1-p_1)r-Ts$}
Next, we minimise $H(q)-H(p)$ over $r$ and $s$, which are general $(d-1)$-dimensional probability vectors.
By (\ref{eq:hqhp}), we have to minimise $H((1-p_1)r-Ts) - H((1-p_1)r)$.
The only extra condition on $r$ and $s$ is $(1-p_1)r-Ts\ge0$.
We will show that minimality is achieved when $(1-p_1)r-Ts$ is rank 1.

\bigskip

\textit{Proof.}
Given that $r$ and $s$ are probability vectors and that the condition $(1-p_1)r-Ts\ge0$
is satisfied, $((1-p_1)r-Ts)/(1-p_1-T)$ is also a probability vector, which we will denote by $\eta$.
Thus $(1-p_1)r-Ts = (1-p_1-T)\eta$.
Conversely, for any pair of probability vectors $s$ and $\eta$, $r':=((1-p_1-T)\eta+Ts)/(1-p_1)$ is a probability vector
satisfying $(1-p_1)r'-Ts\ge0$. Therefore, we can do the substitution $(1-p_1)r-Ts = (1-p_1-T)\eta$ and forget about $r$ altogether.
Thus we are down to minimising
$$
H((1-p_1-T)\eta) - H((1-p_1-T)\eta + Ts)
$$
over all probability vectors $\eta$ and $s$.

Now note that for all $x,y\ge0$, $H(x)-H(x+y)$ is concave and monotonously increasing in $x$. Indeed, the first derivative w.r.t.\ $x$
is $\log(1+y/x)\ge0$ and the second derivative is $-y/(x(x+y))\le0$.
Thus, as in the previous stage, we can conclude that $H((1-p_1-T)\eta) - H((1-p_1-T)\eta + Ts)$ is minimal for an extremal $\eta$.
Since we haven't yet decided on $s$, we will put w.l.o.g.\ $\eta=e^1$.
\qed

With this optimal value for $\eta$, and putting
$$
s=(s_1,(1-s_1)\phi)
$$
(with $\phi$ a $(d-2)$-dimensional probability vector), we get
\beas
\lefteqn{H((1-p_1-T)\eta) - H((1-p_1-T)\eta + Ts)} \\
&=& H(1-p_1-T) - H(1-p_1-T(1-s_1))\\
&& - H(T(1-s_1)\phi)
\eeas
The remaining minimisation over $s$ now consists of first minimising over $\phi$, and then over $s_1$.

The minimisation over $\phi$ is easy, because it only involves the term $H(T(1-s_1)\phi)$, without any constraint
other than that $\phi$ be a probability vector. This term achieves its maximum when $\phi=(1,1,\ldots,1)/(d-2)$, the uniform
distribution, and the maximum value is $T(1-s_1)\log_2(d-2)+H(T(1-s_1))$.

We are now left with a minimisation over $s_1$ of the function
\bea
&&H(1-p_1-T) - H(1-p_1-T(1-s_1)) \nonumber \\
&&- T(1-s_1)\log_2(d-2) - H(T(1-s_1)).\label{eq:hs1}
\eea
We will tackle this minimisation in the next stage.
\subsection{Optimal $s_1$}
In terms of $s_1$, (\ref{eq:hs1}) is the sum of a linear term,
$$
H(1-p_1-T) - T(1-s_1)\log_2(d-2),
$$
and the non-linear term
$$
- H(1-p_1-T(1-s_1)) - H(T(1-s_1)).
$$
This term is of the form $-H(y-x)-H(x)$, with $0\le x\le y$, and is therefore convex in $s_1$.
The only constraint on $s_1$ is that it be in the interval $[0,1]$.

We therefore have find the local minimum of (\ref{eq:hs1}); by convexity of the function, we are guaranteed there is only one.
If this minimum is inside the feasible interval $0\le s_1\le1$, then
this gives the answer; if it is outside it, then the minimum of the constrained minimisation is either 0 or 1, depending
on the location of the local minimum.

The derivative of (\ref{eq:hs1}) w.r.t.\ $s_1$ is
\beas
&& T(\log_2(d-2)+\log_2(1-p_1-T(1-s_1)) \\
&& -\log_2(T(1-s_1))).
\eeas
For $T>0$, this is 0 when
$$
(d-2)(1-p_1-T(1-s_1)) = T(1-s_1),
$$
that is, when
$$
T(1-s_1) = \frac{(d-2)(1-p_1)}{d-1}.
$$
Recall that from the restriction $p_1\le 1-T$ follows $T\le 1-p_1$.
As the LHS lies between 0 and $T$, we have to consider two cases.

\bigskip

\textit{Case (i) --} If $0< T < (d-2)(1-p_1)/(d-1)$, the local optimum cannot be achieved,
and we have to take the nearest point, which is where $T(1-s_1)=T$, i.e.\ $s_1=0$.
Then the minimum of (\ref{eq:hs1}) is given by
\be
- T\log_2(d-2) - H(T).\label{eq:hs1a}
\ee

\textit{Case (ii) --} If $(d-2)(1-p_1)/(d-1)\le T\le 1-p_1$, the local optimum is a feasible point,
and we can put $T(1-s_1) = (d-2)(1-p_1)/(d-1)$.
For the minimum of (\ref{eq:hs1}) this gives
\bea
&&H(1-p_1-T) - H(\frac{1-p_1}{d-1}) \nonumber \\
&&- \frac{(d-2)(1-p_1)}{d-1}\log_2(d-2) \nonumber \\
&&-H(\frac{(d-2)(1-p_1)}{d-1}).\label{eq:hs1b}
\eea

\subsection{Optimal $p_1$}
For the final step of the procedure, we have to find the $p_1$ that minimises the complete expression of the minimum of $H(q)-H(p)$
that we have found so far, under the restriction $0\le p_1\le 1-T$.
We have to consider the two cases from the previous stage.

\textit{Case (i) --} If $T < (d-2)(1-p_1)/(d-1)$, that is, $0\le p_1\le 1-(d-1)T/(d-2)$,
we need to minimise
\be
H(p_1+T)-H(p_1)- T\log_2(d-2) - H(T).
\ee
This case only occurs when $T\le (d-2)/(d-1)$.
By a previously obtained result, the function $x\mapsto H(x+y)-H(x)$ is
monotonously decreasing in $x$ (and convex). Its minimum therefore
occurs for the largest possible value of $p_1$, which in this case is $p_1=1-(d-1)T/(d-2)$.
This gives as minimal value
\bea
&& H(1-T/(d-2)) - H(1-(d-1)T/(d-2)) \nonumber\\
&&- T\log_2(d-2) - H(T).  \label{eq:v1}
\eea

\textit{Case (ii) --} If $(d-2)(1-p_1)/(d-1)\le T\le 1-p_1$, that is,
$1-(d-1)T/(d-2) \le p_1 \le 1-T$,
we need to minimise
\bea
&&H(p_1+T)-H(p_1) \nonumber \\
&&+H(1-p_1-T) - H(\frac{1-p_1}{d-1}) \nonumber \\
&&- \frac{(d-2)(1-p_1)}{d-1}\log_2(d-2) \nonumber \\
&&-H(\frac{(d-2)(1-p_1)}{d-1}). \label{eq:ggg}
\eea
The derivative of (\ref{eq:ggg}) w.r.t.\ $p_1$ equals the logarithm of
$$
\frac{(d-1)p_1(1-p_1-T)}{(1-p_1)(p_1+T)}.
$$
This expression obviously decreases with $T$, and for the minimal allowed value $T=(d-2)(1-p_1)/(d-1)$
it is given by
$$
\frac{(d-2)(1-p_1)}{d-2+p_1},
$$
which is easily seen to be below 1; its logarithm is therefore negative.
Consequentially, the derivative of (\ref{eq:ggg}) is negative over the range under consideration. We conclude that (\ref{eq:ggg})
is minimal for the maximal allowed $p_1$, which is $p_1=1-T$.

This gives as minimal value for $H(q)-H(p)$
\beas
&&H(1)-H(1-T)+H(0) - H(\frac{T}{d-1})\\
&&- \frac{(d-2)T}{d-1}\log_2(d-2)-H(\frac{(d-2)T}{d-1}),
\eeas
which simplifies to
\bea
-(T\log_2(d-1)+H(T)+H(1-T)). \label{eq:v2}
\eea

\bigskip

The final step is now to take the minimum of the two cases (\ref{eq:v1}) and (\ref{eq:v2}), the former one only being valid for
$T\le (d-2)/(d-1)$.
From the fact that $H(x+y)-H(x)$ is monotonously decreasing in $x$ one deduces the relation $H(1-a)+H(1-b)\ge H(1-a-b)$,
for $0\le a,b$ and $a+b\le 1$.
The terms $H(1-T/(d-2)) - H(1-(d-1)T/(d-2))$ in (\ref{eq:v1})
are therefore larger than the term $H(1-T)$ in (\ref{eq:v2}). Furthermore, $- T\log_2(d-2)$
is larger than $- T\log_2(d-1)$. Hence, (\ref{eq:v2}) is always smaller than (\ref{eq:v1}).

Taking absolute values and noting that (\ref{eq:v2}) is always negative then finally
yields inequality (\ref{eq:H}).
\qed
\begin{acknowledgments}
The author thanks the hospitality of the Max Planck Institute for Quantum Optics (in Garching bei M{\"u}nchen)
where this work was initiated. This work
is supported by The Leverhulme Trust (grant F/07 058/U), by EU Integrated Project QAP, and by the Institute of
Mathematical Sciences, Imperial College London, and is part of the QIP-IRC (www.qipirc.org) supported by EPSRC
(GR/S82176/0).
Humble thanks to A.\ Retzker for proofreading.
Thanks also to Norbert Schuch, Michael Wolf and Tobias Osborne for stimulating discussions and for sharing
enjoyment in the beauty of inequality (\ref{eq:sharp}).
\end{acknowledgments}



\begin{thebibliography}{9}
\bibitem{provided} Provided one does not make silly mistakes; thanks to M.\ Reuter for pointing this out to me.
\bibitem{attention} Those readers who proceed through the proof are occasionally rewarded with notoriously bad Flemish jokes.
Believe me, they ain't worth it!
\end{thebibliography}
\end{document}